\documentclass[prd,showkeys,floatfix,nofootinbib,
               preprint,12pt,
               tightenlines,fleqn]{revtex4-1} 

\usepackage{amsmath,amssymb,revsymb,graphicx,dcolumn}
\usepackage{leftidx}
\usepackage{slashed}
\usepackage{marginnote}

\newcommand{\stkout}[1]{\ifmmode\text{\sout{\ensuremath{#1}}}\else\sout{#1}\fi}

\usepackage{xcolor}
\definecolor{mypink}{rgb}{0.9, 0.1, 0.3}

\newcommand{\beq}{\begin{equation}}
\newcommand{\eeq}{\end{equation}}
\newcommand{\beqa}{\begin{eqnarray}}
\newcommand{\eeqa}{\end{eqnarray}}
\newcommand{\bsubeqs}{\begin{subequations}}
\newcommand{\esubeqs}{\end{subequations}}

\begin{document}
\title[]
      {On free fall of quantum matter}
\author{Viacheslav A. Emelyanov}
\email{viacheslav.emelyanov@kit.edu}
\affiliation{Institute for Theoretical Physics,\\
Karlsruhe Institute of Technology,\\
76131 Karlsruhe, Germany\\}

\begin{abstract}
\vspace*{2.5mm}\noindent
We propose an approach that allows to systematically take into account gravity in quantum~particle
physics. It is based on quantum field theory and the general principle of relativity.~These are used~to
build a model for quantum particles in curved spacetime. We compute by its means a deviation
from a classical geodesic in the Earth's gravitational field.~This shows that free fall depends~on~quantum-
matter properties. Specifically, we find that the free-fall universality and~the~wave-packet spreading
are mutually exclusive phenomena.~We then estimate the E\"{o}tv\"{o}s
parameter for a pair of atoms~freely
falling near the Earth's surface, provided that the wave-packet spreading is more
fundamental than the weak equivalence principle.
\end{abstract} 


\maketitle

\section{Introduction}

The theory of quantum fields is well known by now to be extremely successful
in describing scattering processes in elementary particle physics. The very notion
of an elementary particle is based on the Poincar\'{e} group which is in turn the
isometry group of Minkowski spacetime. According~to the general theory of relativity,
Minkowski spacetime describes a universe with no matter~and no cosmological
constant.~The observable Universe is therefore described by~a non-Minkowski
spacetime. The questions arise then how to model a quantum particle in
the absence of the global Poincar\'{e} symmetry and how to experimentally
test~this~construction~in the presence of a gravitational field.

From an experimental point of view, these questions need to be studied in
the~background of the Earth's gravitational field. This may be described
approximately by the Schwarzschild line element (the Earth's rotation neglected)
which, in spherical coordinates,
reads
\beqa\label{eq:schwarzschild}
ds^2 &=& f(r)dt^2 - \frac{dr^2}{f(r)} - r^2\big(d\theta^2 + \sin^2\theta d\phi^2\big)
\quad \text{with} \quad
f(r) \;\equiv\; 1 - \frac{r_S}{r}\,,
\eeqa
where $r_S$ is the Schwarzschild radius, which is $r_{S,\,\oplus} \approx 8.87{\times}10^{-3}\,\text{m}$
in the case of Earth,~while the Earth's radius is $r_{\oplus} \approx 6.37{\times}10^{6}\,\text{m}$.~The
ratio $r_{S,\,\oplus}/r_{\oplus} \approx 1.39{\times}10^{-9}$ is negligibly~small,~yet its gradient
is responsible for the gravitational force, whose gradient is in its turn for the~tidal
effects. These manifestations of the Earth's gravitational field are (for good reason) irrelevant in
particle colliders.~In light of this,~the Schwarzschild line element turns in the~zeroth-order
approximation into
the Minkowski one to underlie the special theory of relativity.

One of the basic postulates of this theory is the special principle of relativity,
which~states that the laws of physics are the same in
all inertial frames~\cite{Einstein-1905}. Even though it was formulated prior to
quantum theory, this principle is implementable in quantum physics. In~particular,~it
gives rise to Wigner's classification of elementary particles~\cite{Wigner-1939}, meaning
that these~are~related to unitary and irreducible representations of the
Poincar\'{e} group, which are distinguished by mass and spin. To put it differently,
the existence of an elementary particle is independent~on an inertial frame
considered.~This mathematical construction is in agreement with up-to-date
observations in high-energy physics.

Collider-physics experiments are performed in the background of the Earth's
gravitational field. For example, non-relativistic neutrons have been shown to
fall down in accordance~with Newton's
gravitational law~\cite{McReynolds,Dabbs&Harvey&Paya&Horstmann,Koester}.
Theoretical particle physics is obviously incomplete,~as~it~is based on the
Minkowski-spacetime approximation and, therefore, the Wigner classification~is
an approximation as well. The fundamental problem is to find a way how to go
beyond~of~this approximation and yet to stay
consistent with collider physics.

From a logical point of view, this should be done by
implementing the general principle~of relativity in quantum theory,
which asserts that physical laws are invariant under general coordinate 
transformations~\cite{Einstein-1916}.
Leaving aside the group-theoretical aspect of this construction, we have recently
shown that the general principle of relativity is implementable~in~de-Sitter
spacetime~\cite{Emelyanov-2020}.~It was gained through deriving a
non-perturbative (in curvature) wave-packet solution which, 
first,~is invariant under the diffeomorphism group~and,~second,~locally reduces 
to the superposition of Minkowski plane waves.~The latter property implies
that~this~solution may be associated locally with
one of the Wigner classes and, therefore, is in agreement with the 
Einstein equivalence principle -- locally and at any point
of the spacetime,~the~Minkowski-spacetime (quantum) physics holds~\cite{Casola&Liberati&Sonego}.

This article aims to treat this approach further.~Specifically, we wish to study the~question if
free fall is universal in quantum theory. In other words, if the weak
equivalence principle~\cite{Casola&Liberati&Sonego} holds in quantum physics.
Furthermore, we shall show in passing that all gravity corrections to
quantum-mechanical phase, which were previously obtained under certain
approximations or heuristically, can be systematically derived by building the
principles~of~general~relativity in quantum particle physics.~We shall find higher-order
gravitational corrections~to~the phase, which, to our knowledge, have not been reported
before, and establish that these corrections may not be unambiguously determined
without experimental data.

Throughout, we use natural units $c = G = \hbar = 1$, unless otherwise stated.

\section{Quantum particles in curved spacetime}

\subsection{Covariant wave packet}
\label{sec:Covariant wave packet}

In order to study quantum corrections to free fall, one needs first to introduce a
model~for quantum particles, since these are elementary ``sensors'' by means of
which one observes free fall in practice. Quantum field theory is by now a
fundamental framework which allows us~to successfully describe microscopic
processes. Its application in particle physics relies, however, on the
Minkowski-spacetime approximation. The outstanding problem remains,~namely
that it is unclear how to consistently take into account gravity in microscopic physics.

We have recently proposed in~\cite{Emelyanov-2020} that an operator, which creates
a quantum scalar particle out of vacuum, should be related to a quantum-field operator,
$\hat{\Phi}(x)$, as follows:
\beqa\label{eq:creation-operator}
\hat{a}^\dagger(\varphi_{X,P}) &\equiv&
\big(\varphi_{X,P},\,\hat{\Phi}\big)_\text{Klein-Gordon}
\nonumber \\[1mm]
&\equiv& - i{\int_\Sigma}d\Sigma^\mu(x)
\left(\varphi_{X,P}(x)\nabla_\mu \hat{\Phi}^\dagger(x) - \hat{\Phi}^\dagger(x)\nabla_\mu\varphi_{X,P}(x)\right),
\eeqa
where $\Sigma$ is a space-like Cauchy surface, $\varphi_{X,P}(x)$ stands for a
wave packet whose centre of mass is localised in the semi-classical limit at
$X = (T,\mathbf{X})$ in coordinate space and at $P = (P^T,\mathbf{P})$ in
momentum space, where the on-shell momentum $P$ belongs to the cotangent
space at $X$. This definition arises from a covariant generalisation of asymptotic
creation and annihilation operators to underlie $S$-matrix in collider
physics~\cite{Lehmann&Symanzik&Zimmermann}.

Accordingly, we obtain from the definition~\eqref{eq:creation-operator} that
\beqa
\big[\hat{a}(\varphi_{X,P}),\,\hat{a}^\dagger(\varphi_{X,P})\big] &=&
\big(\varphi_{X,P},\,\varphi_{X,P}\big)_\text{Klein-Gordon}\,,
\eeqa
where we have made use of the quantum-field algebra (e.g. see~\cite{DeWitt-2003}).~This
commutator~defines the normalisation condition for the wave packet:
\beqa\label{eq:nc}
\big(\varphi_{X,P},\,\varphi_{X,P}\big)_\text{Klein-Gordon} &=& 1\,,
\eeqa
which reduces to the standard normalisation
condition known in quantum mechanics if one considers the
non-relativistic approximation in the weak-field limit, after having rescaled the
wave packet by $1/\sqrt{2M}$, where $M$ is the scalar-field mass.
Thus, a single-particle state is
\beqa
|\varphi_{X,P}\rangle &\equiv& \hat{a}^\dagger(\varphi_{X,P})|\Omega\rangle\,,
\eeqa
where $|\Omega\rangle$ is the quantum vacuum, namely $\hat{a}(\varphi_{X,P})|\Omega\rangle = 0$.
The primary task is, therefore,~to determine $\varphi_{X,P}(x)$ on physical grounds. It should be
emphasised that it~is~by~now~a~standard approach in quantum theory to search instead for a
\emph{global} quantum-field-mode expansion~in~a given curved space~\cite{DeWitt-1975}.
This approach has been put forward as a generalisation~of~the~global plane-wave expansion in
Minkowski spacetime, where quantum field theory serves primarily for the description of scattering
processes~\cite{Weinberg}.~Still, the Universe is not globally flat~\cite{Mukhanov}.~This
means that elementary particle physics is based on the Minkowski-spacetime approximation.
It then follows from the Einstein equivalence principle that the global plane-wave expansion 
of quantum-field operators in particle physics corresponds to a local quantum-field expansion 
in the non-flat Universe. If we now consider, for instance, the Sunyaev-Zeldovich effect, which 
is responsible for the CMB-spectrum distortion due to the inverse Compton scattering of the
CMB photons by high-energy galaxy-cluster electrons, then we conclude that the plane-wave
expansion of quantum-field operators has to locally emerge at any given small-enough~space-time
region. The equivalence
principle suggests further that one needs to deal with an object which depends on a
relative distance, rather than on an absolute position as that is the case for quantum-field
operators, e.g. $\hat{\Phi}(x)$.~The wave function $\varphi_{X,P}(x)$, which
describes~a~particle in quantum theory, seems thereby to be a natural object for that purpose.

The Minkowski-spacetime approximation in elementary particle physics is good enough~to
describe high-energy processes to take place in colliders. Primary observables here are~related
to $S$-matrix elements, each of which provides a probability amplitude for a given
initial~multi-particle state to evolve into a particular final multi-particle state.~Both initial and final
states are usually associated with wave functions to have definite momenta -- plane waves.
However, the LSZ reduction formula~\cite{Lehmann&Symanzik&Zimmermann} (which basically
links the mathematical formalism of quantum field theory to physics) uses normalisable wave
functions to describe particles having neither definite momentum nor position -- superposition of
plane waves. Note, $S$-matrix
has also to depend on initial and final center-of-mass
positions of quantum particles,~otherwise~one could not implement the cluster decomposition
principle, basically stating that distant experiments give uncorrelated results (see Ch.~4
in~\cite{Weinberg}). This requires localised-in-space quantum states which correspond to
wave packets.~We conclude from these trivial observations that $\varphi_{X,P}(x)$ must be
representable via superposition of plane waves for $x$ sufficiently close to $X$, if
treated in a local inertial frame.

The general principle of relativity requires that physical laws are the same in all coordinate
frames~\cite{Einstein-1916}. In particular, the Einstein field equations are invariant under 
general coordinate transformations. In the semi-classical approximation, the single-particle
state $|\varphi_{X,P}\rangle$ enters the Einstein equations through the expectation value
of the stress-tensor operator $\hat{T}_{\mu\nu}(x)$ of the scalar field $\hat{\Phi}(x)$
in the quantum state $|\varphi_{X,P}\rangle$. For this expectation value to be tensorial,
one must impose a condition that $\varphi_{X,P}(x)$ is a covariant wave packet or scalar
in the problem under consideration. Therefore, quantum particle physics must be formulated
in~an~observer- independent manner: All observers, independent on their state of motion or
their rest frame, agree on the existence of the single-particle state $|\varphi_{X,P}\rangle$,
assuming~that~a~quantum~particle, which is described by this state, moves through their
detectors (e.g. Wilson's cloud chambers in case of an electrically charged particle). Still, this
particle affects their detectors differently and this depends on their state of motion (e.g.
curvature of a charged-particle track depends on how a given Wilson chamber moves). This
idea logically follows from the general principle of relativity, but remains
unexplored in quantum particle physics.

By analogy with the Minkowski-spacetime case, we furthermore suppose that $\varphi_{X,P}(x)$ is a
solution of the scalar-field equation, which has the following Fourier-integral representation:
\beqa\label{eq:wp-structure}
\varphi_{X,P}(x) &\equiv& \big({-}g(X)\big)^{-\frac{1}{2}}{\int}\frac{d^4K}{(2\pi)^3}\,\theta(K^T)\,
\delta(K^2-M^2)\,F_P(K)\,\phi_{X,K}(x)\,,
\eeqa
where $\phi_{X,K}(x)$ satisfies the scalar-field equation on mass shell
$g^{AB}(X)K_AK_B \equiv K^2 = M^2$:
\beqa\label{eq:feq}
\Big(\Box_x + M^2 - \frac{1}{6}\,R(x)\Big)\phi_{X,K}(x) &=& 0\,,
\eeqa
and $F_P(K)$ has a peak at $K = P$ and provides for the normalisation
condition~\eqref{eq:nc}.~It~should be remarked at this point that the Klein-Gordon
product, which has been defined in~\eqref{eq:creation-operator},~is~not conserved
if the scalar field interacts with itself or other (non-gravitational) fields.~In~this~case,
$|\varphi_{X,P}\rangle$ is a dynamical state that may, for example, be unstable.
The~normalisation~condition cannot then be fulfilled for all times (e.g. free neutrons
have a mean lifetime of~about~15~min). However, this condition
holds for all times in curved spacetime, even if it is dynamical (e.g. gravitational waves,
collapse etc.) if $(\Box_x + s(x))\hat{\Phi}(x) = 0$ is fulfilled, where $s(x)$ is real.

However, a word of caution is needed on this point. The gravitational interaction between
a pair of particles can be thought as virtual-graviton exchange within the effective
theory~of quantum gravity~\cite{Donoghue}. This quantum-gravity approach presumes a
background gravitational field whose fluctuations are promoted to quantum-field
operators.~Hence, in quantum gravity, $\hat{\Phi}(x)$ satisfies the scalar-field equation~\eqref{eq:feq}
with a non-trivial right-hand side. This results~in~a dynamical evolution of
$\hat{a}(\varphi_{X,P})$.~We treat in this article a test-particle approximation~--~$|\varphi_{X,P}\rangle$
does not source gravity or, in other words, is independent on the graviton field $\hat{h}_{\mu\nu}(x)$.

It proves useful to consider Riemann normal coordinates at $X$, such those $x$
corresponds
to $y$, while $X$ to $Y \equiv (0,\,0,\,0,\,0)$. In the Riemann frame, i.e. $y$, the first derivative of the metric
tensor vanishes at $X$.~Thereby, geodesics passing through the point $X$ turn~into~straight~lines
in the Riemann frame~\cite{Petrov}. Note, this particular choice of coordinates does not affect physics
as we deal with the covariant wave packet. In the Riemann frame, however, the
metric tensor is given through the curvature tensor, its covariant derivatives and their
tensorial products~in a covariant form:
\beqa\nonumber\label{eq:metric-tensor}
g_{ab}(y) &=& \eta_{ab} - \frac{1}{3}\,R_{acbd}\,y^cy^d
- \frac{1}{6}\,R_{acbd;e}\,y^cy^dy^e
\\[1mm]
&&
-\left(\frac{1}{20}\,R_{acbd;ef} -\frac{2}{45}\,R_{acgd}R_{be\;\;f}^{\;\;\;\;g}\right)y^cy^dy^ey^f
+ \text{O}(y^5)\,,
\eeqa
where it is implicitly understood that the curvature tensor and its derivatives are computed
at $X$ (see Sec.~7 in~\cite{Petrov}). The Latin indices running over $\{0,1,2,3\}$ are
coupled to~the~capital Greek indices by means of the vierbein $e_A^a(X)$.~We shall
study in what follows how~these~three curvature corrections to the Minkowski metric
$\eta_{ab}$ affect a quantum particle.

We shall mainly focus here on the wave-packet propagation in the
Schwarzschild geometry in order to find out whether free fall is universal
in quantum theory. We have proposed in~\cite{Emelyanov-2020} to characterise the
wave-packet propagation by its centre-of-mass position:
\beqa\label{eq:position}
\langle y^i \rangle &\equiv& -i{\int_{y^0}}d^3\mathbf{y}\sqrt{-g(y)}\,y^i g^{0b}(y)
\big(\varphi_{Y,P}(y)\partial_b\bar{\varphi}_{Y,P}(y)
- \bar{\varphi}_{Y,P}(y)\partial_b\varphi_{Y,P}(y)\big)\,.
\eeqa
In Minkowski spacetime and in the non-relativistic limit, it reduces to the standard definition of the
position~expectation value $\langle y^i \rangle$, which is known in quantum mechanics, while
$\langle y^0 \rangle = y^0$, i.e. $y^0$ is generically a c-number. Furthermore, the wave-packet
propagation rate then reads
\beqa\label{eq:propagation-rate}
\partial_0\langle y^i \rangle &=& -i{\int_{y^0}}d^3\mathbf{y}\sqrt{-g(y)}\,g^{ib}(y)
\big(\varphi_{Y,P}(y)\partial_b\bar{\varphi}_{Y,P}(y)
- \bar{\varphi}_{Y,P}(y)\partial_b\varphi_{Y,P}(y)\big)\,,
\eeqa
where we have used the normalisation condition~\eqref{eq:nc}, the scalar-field
equation~\eqref{eq:feq} re-written in Riemann normal coordinates and also the fact that
$\varphi_{Y,P}(y)$ vanishes at spatial infinity~due~to its localisation in space.~It
is straightforward to show that this rate reduces~to~the momentum expectation value
divided by $M$ in the quantum-mechanics regime. For all these reasons,~$\langle y^i \rangle$ defined
in~\eqref{eq:position} might be a proper starting point to study free fall of quantum matter.

\subsection{Non-inertial effects}

A physically relevant solution $\phi_{X,K}(x)$ of the scalar-field equation~\eqref{eq:feq}
must be determined from observations in particle physics. To our knowledge, there
has been made~up~to~now~only one observation which reveals the role of the
curvature tensor in quantum particle physics~and which we shall utilise below. We intend first
to neglect the space-time curvature to~obtain~the zeroth-order term (in curvature) of the wave packet
$\varphi_{X,P}(x)$ in curved spacetime.

In the absence of the space-time curvature, we obtain from~\eqref{eq:feq} that
\beqa\label{eq:wps-rc-lot}
\phi_{Y,K}^{(0)}(y) &=& e^{-iK{\cdot}y}\,,
\eeqa
where
\beqa
K{\cdot}y & \equiv & K_a y^a\,,
\eeqa
as this directly follows from quantum field theory in Minkowski spacetime and its application to
elementary particle physics~\cite{Weinberg}.~Next, we obtain a covariant Gaussian wave
packet~\cite{Naumov&Naumov,Naumov}, with the momentum variance $D > 0$, namely
\beqa\label{eq:app-wp-0}
\varphi_{Y,P}^{(0)}(y) &=& N^{(0)}\,
\frac{K_1\big(\frac{M^2}{D^2}\Sigma_{Y,P}(y)\big)}{\Sigma_{Y,P}(y)}\,,
\eeqa
where $K_\nu(z)$ is the modified Bessel function of the second kind,
\bsubeqs
\beqa
\Sigma_{Y,P}(y) &\equiv&
\left(\frac{1}{4}
+i\,\frac{D^2}{M^2}\,P{\cdot}y - \frac{D^4}{M^2}\,y{\cdot}y\right)^\frac{1}{2}
\eeqa
and
\beqa\label{eq:normalisation-factor-0}
N^{(0)} &\equiv& \frac{D}{2\pi\sqrt{K_1\hspace{-1.0mm}\left(\frac{M^2}{D^2}\right)}}\,.
\eeqa
\esubeqs
We refer to Sec.~B in~\cite{Emelyanov-2020}, where we have computed the center-of-mass
propagation of this~wave packet and its energy-momentum vector, and found that this packet
behaves kinematically~as a point-like particle of the same mass if and only if $Mc \gg D$.~This
may thereby be interpreted as a classical limit in which particles can be treated
as effectively having no extent.

It is worth mentioning that the right-hand side of~\eqref{eq:wps-rc-lot}~and, consequently,
the right-hand side of~\eqref{eq:app-wp-0} can be expressed in terms of general
coordinates, namely
\beqa\label{eq:wps-gc-lot}
\phi_{X,K}^{(0)}(x) &=& e^{+iK{\cdot}\sigma}\,,
\eeqa 
where
\beqa
K{\cdot}\sigma &\equiv& K_M\,g^{MN}(X)\,\nabla_N\sigma(x,X)
\eeqa
and $\sigma \equiv \sigma(x,X)$ is the geodetic distance~\cite{DeWitt-1965}. The capital
indices refer to the tangent space at $X$. The zeroth-order solution of the scalar-field
equation in the form~\eqref{eq:wps-gc-lot} was our starting point in~\cite{Emelyanov-2020}
to obtain a covariant wave-packet solution in de-Sitter spacetime, which, first, is
non-perturbative in curvature and, second, reduces locally to the plane-wave~superposition
at any point $X$. The latter property implies that this solution is consistent with the application
of quantum field theory to collider physics.

The approximate solution~\eqref{eq:wps-rc-lot} (or, alternatively,~\eqref{eq:wps-gc-lot}) can
be used to make ``predictions" which can then be compared with observations in quantum
particle physics by making use~of the relation between the Schwarzschild and Riemann
normal coordinates.

\subsubsection{Wave-packet propagation}

In Riemann normal coordinates, all geodesics which pass through the point $X$ are
given~by a straight line~\cite{Petrov}. For instance, we consider the following (classical)
geodesic:
\beqa\label{eq:classical-geodesic}
y^a(\tau) &=& (P^a/M)\,\tau\,,
\eeqa
where $\tau$ is the proper time. This geodesic can be re-written in terms of general
coordinates, namely
\beqa\label{eq:classical-geodesic-general}
x^a(\tau) &=& x^a(y)\big|_{y \,=\, (P/M)\tau}\,.
\eeqa
This is the result of classical theory. In quantum theory, one should instead consider
\beqa\label{eq:quantum-geodesic}
\langle x^a \rangle & \equiv & -i{\int_{y^0}}d^3\mathbf{y}\sqrt{-g(y)}\,x^a(y)\,g^{0b}(y)
\big(\varphi_{Y,P}(y)\partial_b\bar{\varphi}_{Y,P}(y)
- \bar{\varphi}_{Y,P}(y)\partial_b\varphi_{Y,P}(y)\big)\,.
\eeqa
This integral cannot in general be evaluated exactly.~We shall do this perturbatively
in~terms of number of metric-tensor derivatives. 

Having expanded $g_{ab}(y)$ (and $\varphi_{Y,P}(y)$) in~\eqref{eq:quantum-geodesic}
over the curvature tensor, its derivatives~and products, and then collected terms having the same
number of the metric derivatives, we~get
\beqa\label{eq:quantum-geodesic-curvature-expansion}
\langle x^a \rangle & = & \langle x^a \rangle_{(0)} +  \sum\limits_{s \,=\, 2}^{\infty}\,\langle x^a \rangle_{(s)}\,,
\eeqa
where the term with $s = 1$ is absent, since we work in Riemann normal
coordinates in which the first derivative of the metric tensor vanishes at $X$. We have, by definition, for
$s = 0$ that
\beqa\label{eq:rnc-trajectory-0}
\langle x^a \rangle_{(0)} & \equiv & 
-i{\int_{y^0}}d^3\mathbf{y}\,x^a(y)\,
\varphi_{Y,P}^{(0)}(y)\partial_0\bar{\varphi}_{Y,P}^{(0)}(y) +\text{c.c.}
\;=\; x^a + \sum\limits_{n \,=\, 0}^{\infty}\delta_{(0)}^{(n)} x^a\,.
\eeqa
Next, in the non-relativistic approximation, i.e. the group velocity $V^i \equiv P^i/P^0 $ is negligible
with respect to the speed of light $c$, we find at $X = (0,\,0,\,0,\,r_\oplus)$ that
\bsubeqs
\beqa
\delta_{(0)}^{(0)}x^a & \xrightarrow[V \,\rightarrow\, 0]{} & 0\,,
\\[1mm]
\delta_{(0)}^{(1)}x^a & \xrightarrow[V \,\rightarrow\, 0]{} &
-\,\frac{g_\oplus\hbar^2}{8D^2c^2}\,\delta_3^a
\left(\Delta_{(0),\,0}^{(1)}+\frac{4D^4}{M^2\hbar^2}\,\Delta_{(0),\,2}^{(1)}\,\tau^2\right),
\\[1mm]
\delta_{(0)}^{(2)}x^a & \xrightarrow[V \,\rightarrow\, 0]{} &
+\,\frac{(g_\oplus)^2\hbar^2}{4D^2c^4}\,\delta_0^a
\left(\Delta_{(0),\,0}^{(2)}+\frac{4D^4}{M^2\hbar^2}\,\Delta_{(0),\,2}^{(2)}\,\tau^2\right)
c\tau\,,
\\[1mm]\label{eq:for-estimate}
\delta_{(0)}^{(3)}x^a & \xrightarrow[V \,\rightarrow\, 0]{} &
+\,\frac{(g_\oplus)^2\hbar^4}{24D^4c^4r_\oplus}\,\delta_3^a
\left(\Delta_{(0),\,0}^{(3)}+\frac{D^2c^2}{\hbar^2}\,\Delta_{(0),\,2}^{(3)}\,\tau^2
+\frac{4D^6c^2}{M^2\hbar^4}\,\Delta_{(0),\,4}^{(3)}\,\tau^4\right),
\\[1mm]
\delta_{(0)}^{(4)}x^a & \xrightarrow[V \,\rightarrow\, 0]{} &
+\,\frac{(g_\oplus)^2\hbar^4}{10D^4c^4(r_\oplus)^2}\,\delta_0^a
\left(\Delta_{(0),\,0}^{(4)}+\frac{D^2c^2}{\hbar^2}\,\Delta_{(0),\,2}^{(4)}\,\tau^2
+\frac{4D^6c^2}{M^2\hbar^4}\,\Delta_{(0),\,4}^{(4)}\,\tau^4\right)
c\tau\,,
\eeqa
\esubeqs
where we have replaced $y^0$ by $\tau$ in accordance with~\eqref{eq:classical-geodesic}
and the non-relativistic limit, 
\beqa
\Delta_{(s),\,k}^{(n)} &\equiv& 1 + \sum\limits_{l \,=\,1}^\infty\,C_{(s),\,k,\,l}^{(n)}\Big(\frac{D}{Mc}\Big)^{2l}\,,
\eeqa
where $C_{(s),\,k,\,l}^{(n)}$ are numerical coefficients, and
\beqa
g_{\oplus} &\equiv& \frac{c^2r_{S,\,\oplus}}{2(r_\oplus)^2} \;\approx\; 9.81\,\text{m/s}^2
\eeqa
is the free-fall acceleration. If we suppose for the moment that $\tau \ll M\hbar/2D^2$, then
all~these corrections to the classical
geodesic~\eqref{eq:classical-geodesic-general} are owing to $D < \infty$.~This
can be readily understood by looking at the argument of the modified Bessel
function in~\eqref{eq:app-wp-0}. Specifically, we obtain~in the limit $M/D \rightarrow \infty$
that
\beqa
\frac{M^2}{D^2}\,\Sigma_{Y,P}(y) & \approx & \frac{M^2}{2D^2}
+ iP{\cdot}y + \left((P{\cdot}y)^2-M^2y{\cdot}y\right)\frac{D^2}{M^2}\,,
\eeqa
where the first (divergent) term is canceled in~\eqref{eq:app-wp-0} by an analogous
term in~\eqref{eq:normalisation-factor-0}, the second term corresponds to the
quantum-mechanical phase, and the last term ensures that the wave packet is
suppressed away from the classical
geodesic~\eqref{eq:classical-geodesic}.~The~strength~of~this~suppression~is
characterised by the momentum variance $D$.~The bigger its value, the smaller
the wave-packet localisation region and, therefore, the deviation from the classical
geodesic must disappear~in the limit $D \rightarrow \infty$, but $D \ll Mc < \infty$
holds in practice.~However,~the~wave-packet spreading starts to play a role if
$\tau \gtrsim M\hbar/2D^2$ (see Sec.~4 of Ch.~2 in~\cite{Merzbacher}). Hence, we
find that~the~wave-packet spreading enhances the deviation from the
geodesic over~a~long-enough time.~Does~this lead to a measurable effect?

Up to the first order in derivatives of the metric tensor, we get
in the~non-relativistic and weak-field limit near the Earth's surface that
\bsubeqs\label{eq:newton-trajectory-0}
\beqa
\langle x^0(\tau) \rangle &\approx& c \tau\,,
\\[1mm]
\langle x^i(\tau) \rangle &\approx& x^i(0) + V^i\tau
- \frac{g_{\oplus}\delta_3^i}{2}
\left(\left(1+\frac{D^2}{M^2c^2}\,\Delta_{(0),\,2}^{(1)}\right)\tau^2
+ \frac{\hbar^2}{4D^2c^2}\,\Delta_{(0),\,0}^{(1)}\right),
\eeqa
\esubeqs
where the wave-packet spreading yields a contribution which is in fact negligible in
the~semi-classical approximation $Mc/D \ggg 1$.~Thus, this trajectory agrees with
experimental tests~of free fall of non-relativistic neutrons nearby the Earth's
surface~\cite{McReynolds,Dabbs&Harvey&Paya&Horstmann,Koester}, where
the spin degree~of freedom does not play any role here, if the semi-classical
approximation applies. If not, then the centre of mass
of the packet propagates towards Earth with a constant acceleration that
differs from $g_\oplus$ by a factor of $1 + (D/Mc)^2$.

Higher-order (in metric derivatives) corrections to the centre-of-mass
trajectory may start to play a role over long-enough time intervals, as, for instance, it follows
from~\eqref{eq:for-estimate}.~However, we need to take into account the
curvature tensor,~because the space-time curvature may~give a non-zero
contribution at that order. We shall study this later below.

\subsubsection{Wave-packet phase}
\label{sec:wpp}

Thermal neutrons have been shown to acquire a phase shift as a consequence of the Earth's
gravitational field~\cite{Colella&Overhauser&Werner}. 
The quantum interference induced by gravity was originally predicted from the Schr\"{o}dinger
equation with the Newtonian gravitational potential~\cite{Colella&Overhauser} and this~result
immediately follows in the non-relativistic limit from the approximate solution~\eqref{eq:app-wp-0}:
\bsubeqs
\beqa\label{eq:phase-0}
\varphi_{X,P}(x)
&\propto& \exp\left(-iMc^2t\left(1 + v^{(1)}\right)\right),
\eeqa
where by definition
\beqa\label{eq:phase-v1}
\delta^{(1)}v &\equiv& \frac{g_\oplus z}{c^2}\,,
\eeqa
\esubeqs
where $z$ is the vertical height to vanish at the Earth's surface.

A few remarks are in order. First, this result has been computed by
considering isotropic coordinates, $(ct,x,y,z)$, in Schwarzschild spacetime, such that
$X = (0,0,0,r_\oplus)$.~We~get~from the coordinate transformation $y = y(x)$
treated at $X$ that
\beqa\label{eq:riemann-coordinates}
y^a &=& x^a + \sum\limits_{n \,=\, 1}^{\infty}\delta^{(n)}y^a\,,
\eeqa
where $x$ vanishes from now on at $X$ and, to the first order in derivatives of the metric tensor,
\beqa\label{eq:riemann-time-1}
\delta^{(1)}y^0 &\approx& +\frac{g_\oplus z}{c^2}\,t\,,
\eeqa
where the approximate equality means that we have omitted polynomials of
the ratio $r_{S,\,\oplus}/r_\oplus$ in the prefactor of the right-hand side. We shall tacitly do
the same below in the higher-order terms in the expansion~\eqref{eq:riemann-coordinates}.
Second, the on-mass-shell condition $P{\cdot}P = M^2c^4$ gives
\beqa
P_T & \xrightarrow[V \,\rightarrow\, 0]{} & \sqrt{g_{TT}(X)}\,Mc^2 \;\approx\;
\left(1-\frac{r_{S,\,\oplus}}{2r_\oplus}\right)Mc^2\,.
\eeqa
This result can be derived by taking into account that $P{\cdot}y = Mc^2\tau$ on the
classical~geodesic, where, in the non-relativistic approximation, the proper time $\tau$ reads
\beqa
\tau &=& \sqrt{g_{AB}(X)y^Ay^B} \;\approx\; \sqrt{g_{TT}(X)}\,y^T\,.
\eeqa
The ratio $r_{S,\,\oplus}/r_\oplus$ in $g_{TT}(X)$ has been neglected in~\eqref{eq:phase-0} as well.

Another experiment which probes the wave-packet-phase structure has been performed~by
making use of an accelerated interferometer~\cite{Bonse&Wroblewski}. Namely, 
the acceleration-induced quantum interference of non-relativistic neutrons has been observed,
that agrees with the expectations from the Einstein principle~\cite{Nauenberg}.
Expressing Riemann normal coordinates~$y$~through~Rindler coordinates $(t_R,x_R,y_R,z_R)$, i.e.
\bsubeqs
\beqa
y^0 &=& \big(c/a + z_R/c\big)\sinh(at_R/c)\,,
\\[1mm]
y^1 &=& x_R\,,
\\[1mm]
y^2 &=& y_R\,,
\\[1mm]
y^3 &=& \big(c^2/a + z_R\big)\cosh(at_R/c)\,,
\eeqa
\esubeqs
we obtain
\beqa\label{eq:phase-rindler}
P{\cdot}y
&\approx& Mc^2t_R\left(1 + \frac{a z_R}{c^2} + \frac{(at_R)^2}{6 c^2}\right),
\eeqa
which coincides in the non-relativistic approximation with the gravity-induced phase shift~if
we take $a = g_\oplus$. The acceleration-squared term cannot be compared yet with
that in gravity, as we first need to take into account higher-order gravity corrections
to~\eqref{eq:phase-0}. We~shall~study these corrections shortly.

The Colella-Overhauser-Werner experiment~\cite{Colella&Overhauser&Werner}
shows that the wave function of a freely-falling particle is a superposition of the scalar-field modes which
are not eigenfunctions of~the Schwarzschild-time translation operator $\partial_t$, even though it is a
Killing vector.~For~this~reason, quantum-field modes, which are eigenfunctions of a time-like Killing
vector, do not necessary correspond to modes whose superposition can be related to a physical
quantum particle.~The Bonse-Wroblewski experiment~\cite{Bonse&Wroblewski} gives another
example for this observation, now in case of Rindler spacetime. However, this circumstance might change if
one considers interacting field models.~For instance,~stationary wave functions above a reflecting plate
describe bound~states of particles in the Earth's gravitational field~\cite{Luschikov&Frank}, which were 
observed in Nature~\cite{Nesvizhevsky&etal}.

The gravity-induced quantum interference has been observed so far in the non-relativistic
regime. The wave-packet phase $P{\cdot}y$ turns on shell into $M\tau$, where
$\tau$ is the proper time over a geodesic defined by the initial centre-of-mass position
$X$~and~the initial momentum~$P$.~Hence, a general relativistic result for the phase
difference in the Colella-Overhauser setup~\cite{Colella&Overhauser} (with $M > 0$ and
the curvature tensor neglected) reads
\beqa
\delta\varphi(h) &=& \big(Mc^2/\hbar\big)\big(\tau_{r_\oplus} - \tau_{r_\oplus+h}\big)\,,
\eeqa
where $h$ is a vertical height of the upper horizontal path above the Earth's surface.
Moreover, we obtain from $P{\cdot}y$ up to the first order in $g_\oplus$ that
\beqa\label{eq:phase-shift}
\delta\varphi(h) &\approx& -\frac{g_\oplus hl}{\hbar c^2}\,\frac{(Mc)^2+2P^2}{P}\,,
\eeqa
where $l$ is the length of the horizontal path. The gravity-induced phase shift $\delta\varphi(h)$
reduces to the Colella-Overhauser result~\cite{Colella&Overhauser} in the non-relativistic
regime, $Mc \gg P$. In the relativistic limit, $\delta\varphi(h)$ is by a factor of two
bigger than that previously reported in~\cite{Mannheim,Zych&etal-2021,Rideout&etal-2021}
for photons. Note, we take in~\eqref{eq:phase-shift} into account both the gravitational time
dilation and length contraction in the horizontal direction, which non-negligibly contributes
in the relativistic limit.~This~was also taken into consideration earlier in~\cite{Brodutch&etal}.

\subsubsection{Preliminary conclusion}

There are several gravity-attributed effects which were experimentally confirmed in non-relativistic
quantum physics. In full agreement with~\cite{McReynolds,Dabbs&Harvey&Paya&Horstmann,Koester},
we have found that the relativistic wave packet is consistent with Ehrenfest's theorem for free fall:
\beqa\label{eq:cf-1}
\frac{d^2}{d\tau^2}\,\langle x^i(\tau) \rangle &\approx& - g_\oplus
\left(1+\frac{D^2}{M^2c^2}\right)\delta_3^i\,,
\eeqa
iff $Mc \ggg D$ is fulfilled.~Besides, thermal neutrons have been shown to non-trivially
interfere with each other due to the Earth's gravitational
field~\cite{Colella&Overhauser&Werner}.~This was predicted earlier
in~\cite{Colella&Overhauser}~by relying on the Schr\"{o}dinger
equation with Newton's potential. Here the covariant wave packet gives the phase
shift~\eqref{eq:phase-shift} which also agrees with the observations. Note that both
effects~are essentially due to the gravitational time
dilation~\eqref{eq:riemann-time-1} (see also~\cite{Czrnecka&Czarnecki}).

For these reasons, the mathematical model for an elementary particle in curved spacetime,
that has been presented in Sec.~\ref{sec:Covariant wave packet}, deserves further
scrutiny.~The next step we intend~now~to make is to perturbatively take into account the
space-time curvature.

\subsection{Curvature effects}

The curvature-dependent terms neglected in the previous section may start to play a
role over long-enough time intervals even if the wave-packet localisation size is much
smaller~than the local curvature length.~In order to study this issue, we look for the
wave-packet solution~in the following form:
\beqa
\phi_{Y,K}(y) &=& \phi_{Y,K}^{(0)}(y) + \sum\limits_{n \;=\; 2}^{\infty}\phi_{Y,K}^{(n)}(y)\,.
\eeqa
Substituting $\phi_{Y,K}(y)$ into the scalar-field equation~\eqref{eq:feq}, we obtain
\beqa\label{eq:wps-rc-lot-n}
\big(\eta^{ab}\partial_a\partial_b + M^2\big)\phi_{Y,K}^{(n)}(y) &=& j^{(n)}(y)\,,
\eeqa
where we have for $n \in \{2,\,3,\,4\}$ that
\bsubeqs
\beqa
j^{(2)}(y) &=& - \mathcal{D}^{(2)}\phi_{Y,K}^{(0)}\,,
\\[3mm]
j^{(3)}(y) &=& - \mathcal{D}^{(3)}\phi_{Y,K}^{(0)}\,,
\\[3mm]
j^{(4)}(y) &=& -\mathcal{D}^{(4)}\phi_{Y,K}^{(0)} - \mathcal{D}^{(2)}\phi_{Y,K}^{(2)}\,,
\eeqa
\esubeqs
and, in vacuum ($R_{ab} = 0$ and, consequently, $\eta^{ab}R_{ab} = 0$),
\bsubeqs
\beqa
\mathcal{D}^{(2)} &=& \frac{1}{3}\,R_{\;\;c\;\;d}^{a\;\;b}\,y^cy^d\partial_a\partial_b\,,
\\[1mm]
\mathcal{D}^{(3)} &=&\frac{1}{6}\,R_{\;\;c\;\;d;e}^{a\;\;b}\,y^cy^dy^e\partial_a\partial_b\,,
\\[1mm]
\mathcal{D}^{(4)} &=&\left(\frac{1}{20}\,R_{\;\;c\;\;d;ef}^{a\;\;b}
+\frac{1}{15}\,R_{\;\;cgd}^a R_{\;\;e\;\;f}^{b\;\;g}\right)y^cy^dy^ey^f
\partial_a\partial_b
-\frac{4}{45}\,R_{\;\;\;\;\;\;a}^{fde}R_{fbec}\,y^ay^by^c\partial_d\,.
\eeqa
\esubeqs
Note that there is no correction with $n = 1$, since we work here in local inertial coordinates.
We now intend to solve this system of differential equations in order.

\subsubsection{LO curvature correction}

We obtain one curvature-dependent term at the leading order (LO) of perturbation~theory,
which does not vanish in vacuum, namely
\beqa\label{eq:wps-rc-lot-1}
\phi_{Y,K}^{(2)}(y) &=& e^{-iK{\cdot}y}\,R_{acbd}K^aK^by^cy^d\,v_1\,,
\eeqa
where $v_1$ is a covariant function of $y$ and $K$. In other words, $v_1$ depends
on $y{\cdot}y \equiv y^2$ and~$K{\cdot}y$ only. We obtain then
from~\eqref{eq:wps-rc-lot-n} with $n=2$ that
\bsubeqs
\beqa\label{eq:wps-rc-lot-1-v}
\partial{\cdot}\partial v_1 - 2iK{\cdot}\partial v_1 + 8\dot{v}_1 &=& \frac{1}{3}\,,
\eeqa
where the dot stands for the differentiation with respect to $y^2$, and the
first two terms on~the left-hand side of this equation can be re-written in terms
of $y^2$ and $K{\cdot}y$ and derivatives of~$v_1$ with respect
to its variables. This equation has infinitely many solutions.~However,~bearing~in mind
$j^{(2)}(y) = \text{O}(y^2)$, one should have
$v_1 \propto K{\cdot}y$. This gives from~\eqref{eq:wps-rc-lot-1-v} that
\beqa\label{eq:wps-rc-lot-1-v-sol}
v_1 &=& \frac{i}{6M^2}\,K{\cdot}y\,.
\eeqa
\esubeqs
It is worth emphasising that this solution is non-unique. For example, $v_1 + \text{const}$ is
another solution of~\eqref{eq:wps-rc-lot-1-v}. This constant represents a free parameter.
It cannot be determined~even~in de-Sitter spacetime even if one considers the de-Sitter
quantum state. It seems, thereby,~that we need experimental data to deal with this
mathematical ambiguity.

There has been recently observed a wave-packet phase shift due to the
curvature~\cite{Asenbaum&etal}.~This effect was predicted earlier by assuming the
non-relativistic approximation~\cite{Anandan,Audretsch&Marzlin-1994-I,
Audretsch&Marzlin-1994-II,Audretsch&Marzlin-1996,Bongs&etal}. In order to show how this result
follows from the covariant wave packet, we first establish the leading-order correction
to~\eqref{eq:app-wp-0}:
\beqa\label{eq:app-wp-1}
\varphi_{Y,P}^{(2)}(y) &=& \frac{1}{48M^2}\,N^{(0)}R_{acbd}\,P^aP^by^cy^d\,
\big(iP{\cdot}y-2D^2y^2\big)\,
\frac{K_4\big(\frac{M^2}{D^2}\Sigma_{Y,P}(y)\big)}{(\Sigma_{Y,P}(y))^4}\,.
\eeqa
Second, choosing isotropic coordinates in Schwarzschild spacetime and
assuming $Mc \gg D$, we find in the non-relativistic approximation up to the second order in
the metric derivatives that
\bsubeqs
\beqa\label{eq:phase-1}
\varphi_{X,P}(x)
&\propto& \exp\left(-iMc^2t\left(1 + v^{(1)} + v^{(2)}\right)\right),
\eeqa
where $\delta^{(1)}v$ has been defined in~\eqref{eq:phase-v1} and
\beqa\label{eq:phase-v2}
v^{(2)} &\equiv& -\frac{1}{2}\,R_{0i0j}\,x^ix^j+\frac{(g_\oplus t)^2}{6c^2}\,,
\eeqa
\esubeqs
where we have also used the expansion of Riemann coordinates via
isotropic coordinates~up~to the second order in derivatives of the metric, appearing
in~\eqref{eq:riemann-coordinates}. The numerical~factor in front of the
curvature-dependent term~in~the phase correction~\eqref{eq:phase-v2} follows
from the combination of the proper-time expansion over~the~non-inertial
coordinates $x$ at $X$ and the leading-order correction
\eqref{eq:app-wp-1} to \eqref{eq:app-wp-0}. This coefficient can also be obtained
by working in Fermi coordinates and expressing the proper time~$\tau$ through the
Fermi time coordinate (see~(45) in~\cite{Li&Ni}). This result agrees with the
observation~\cite{Asenbaum&etal} and, therefore, the
imaginary part of $v_1$ given in~\eqref{eq:wps-rc-lot-1-v-sol} is physically correct.

The acceleration-squared term in~\eqref{eq:phase-v2} remains experimentally unobserved, 
although,~as~a matter of principle, it may be testable, as recently argued
in~\cite{Marletto&Vedral}.~This term~manifests~itself~in the non-inertial 
frame and is, accordingly, present in~\eqref{eq:phase-rindler} as well (see also~\cite{Nauenberg}).

The curvature-dependent correction~\eqref{eq:app-wp-1} is covariant. Besides, it
vanishes on the classical geodesic~\eqref{eq:classical-geodesic}. Still, the wave
packet has a non-vanishing support in space. This correction is, hence, generically
non-zero iff $D < \infty$ (note that the packet is proportional to~the~Wightman
function in the limit $D \rightarrow \infty$, which, in vacuum, is oblivious to the
leading-order curvature correction, see~(2.21) in~\cite{Bunch&Parker}). Specifically,
$\varphi_{X,P}(x)$ becomes deformed due to~\eqref{eq:app-wp-1}. Moreover, this
deformation is typical for the gravitational tidal effect, namely the wave packet
becomes squeezed in the horizontal direction, while stretched in the vertical one.

For this reason, one might expect that~\eqref{eq:app-wp-1} influences the wave-packet
propagation solely due to $D < \infty$. Indeed, we have
from~\eqref{eq:quantum-geodesic-curvature-expansion} that
\beqa\label{eq:rnc-trajectory-2}
\langle x^a \rangle_{(2)} & \equiv & \sum\limits_{n \,=\, 0}^{\infty}\delta_{(2)}^{(n)} x^a\,,
\eeqa
where, up to the fourth order in derivatives of the metric tensor, we find
in the non-relativistic limit in Schwarzschild spacetime that
\bsubeqs
\beqa
\delta_{(2)}^{(0)}x^a & \xrightarrow[V \,\rightarrow\, 0]{} & 0\,,
\\[2mm]
\delta_{(2)}^{(1)}x^a & \xrightarrow[V \,\rightarrow\, 0]{} &
+\,\frac{(g_\oplus)^2 \hbar^4}{3D^4c^4r_\oplus}\,\delta_3^a
\left(\Delta_{(2),\,2}^{(1)}+\frac{D^2c^2}{\hbar^2}\,\Delta_{(2),\,2}^{(1)}\,\tau^2\right),
\\[2mm]
\delta_{(2)}^{(2)}x^a & \xrightarrow[V \,\rightarrow\, 0]{} & 0\,.
\eeqa
\esubeqs
It should be noted that there is no contribution to the normalisation
factor~\eqref{eq:normalisation-factor-0} at this~order of perturbation theory.
In fact, both the wave packet and its normalisaition condition~\eqref{eq:nc}~do
not depend on a coordinate frame. This means that the leading-order correction
to~\eqref{eq:normalisation-factor-0}~has to be proportional to the Riemann tensor
$R_{abcd}$ contracted with the metric tensor $\eta_{ab}$ and~the initial four-momentum
$P_a$, as there are no other tensors in the problem under consideration.
Yet, any scalar obtained by contracting $R_{abcd}$ with $\eta_{ab}$ and $P_a$ is zero in vacuum.

Up to the second order in derivatives of the metric tensor, we obtain
in the non-relativistic limit at the Earth's surface for
$\tau \ll c/g_\oplus \approx 3.06{\times}10^{7}\,\text{s}$, i.e. $\tau \approx t$, that
\bsubeqs\label{eq:newton-trajectory-2}
\beqa
\langle x^0(\tau) \rangle &\approx& c\tau
\left(1+\frac{(g_\oplus)^2\hbar^2}{4D^2c^4}\,\Delta_{(0),\,0}^{(2)}\right),
\\[1mm]
\langle x^i(\tau) \rangle &\approx& x^i(0) + V^i \tau
- \frac{g_{\oplus}\delta_3^i}{2}
\left(\left(1+\frac{D^2}{M^2c^2}\,\Delta_{(0),\,2}^{(1)}\right)\tau^2
+ \frac{\hbar^2}{4D^2c^2}\,\Delta_{(0),\,0}^{(1)}\right).
\eeqa
\esubeqs
The novel contribution entering the centre-of-mass trajectory of the wave packet
at this~order influences time duration of free fall (cf.~\cite{Viola&Onofrio}). 
This result shows that the position expectation values defined with respect to $\tau = \text{const}$
and $\langle x^0(\tau) \rangle = \text{const}$ Cauchy surfaces approximately agree if
$\hbar/D \ll r_\oplus(r_\oplus/r_{S,\,\oplus}) \sim 100{\times}\text{solar-system size}$.

\subsubsection{NLO curvature correction}

There are two curvature-dependent terms in the next-to-leading order (NLO):
\beqa\label{eq:wps-rc-lot-2}
\phi_{Y,K}^{(3)}(y) &=& e^{-iK{\cdot}y}
R_{acbd;e}K^aK^by^cy^d\left(w_1y^e + w_2K^e\right)\,,
\eeqa
where $w_1$ and $w_2$ are covariant functions of $y$ and $K$, which do not
vanish in vacuum. First, we obtain from~\eqref{eq:wps-rc-lot-n} with $n=3$ that
\bsubeqs
\beqa\label{eq:wps-rc-lot-2-w-1}
\partial{\cdot}\partial w_1 - 2iK{\cdot}\partial w_1 + 12\dot{w}_1 &=& \frac{1}{6}\,.
\eeqa
According to the minimal Ansatz, we need to look for a solution of the
inhomogeneous~part~of this equation only. By analogy with our procedure
in the previous section, we obtain
\beqa
w_1 &=& \frac{i}{12M^2}\,K{\cdot}y\,.
\eeqa
\esubeqs
Second, we have
\bsubeqs
\beqa\label{eq:wps-rc-lot-2-w-2}
\partial{\cdot}\partial w_2 - 2iK{\cdot}\partial w_2 + 8\dot{w}_2 &=& 2iw_1 - 2w_1^\prime\,,
\eeqa
where the prime stands for the differentiation with respect to $K{\cdot}y$.
The inhomogeneity~of~this equation originates from $w_1 \propto K{\cdot}y$.
Therefore, we assume
that $w_2$ depends on $K{\cdot}y$ only.~This results in
\beqa
w_2 &=&-\frac{i}{24M^4}\,(K{\cdot}y)^2 + \frac{1}{24M^4}\,K{\cdot}y\,,
\eeqa
\esubeqs
which has no free parameters.

Even though there are no experimental data, to our knowledge, which need
the curvature derivative for their explanation, we can make a prediction by using
this particular solution~in the non-relativistic regime. This is due to
$R_{0c0d;0} = 0$
in Schwarzschild~spacetime.~Therefore, the wave-packet phase up to the third order in
derivatives of the metric tensor reads
\bsubeqs
\beqa
\varphi_{X,P}(x)
&\propto& \exp\left(-iMc^2t\,\left(1 + v^{(1)} + v^{(2)} + v^{(3)}\right)\right),
\eeqa
where $v^{(1)}$ and $v^{(2)}$ have, respectively, been defined in~\eqref{eq:phase-v1}
and~\eqref{eq:phase-v2}, and
\beqa
v^{(3)} &\equiv& -\frac{1}{6}\,R_{0i0j;k}\,x^ix^jx^k -\frac{2(g_\oplus t)^2}{3c^2}\,\frac{z}{r_\oplus}\,,
\eeqa
\esubeqs
and we have also made use of~\eqref{eq:riemann-coordinates} up to the order with
$n = 3$.~The numerical factor in~front~of the curvature-dependent term coincides
with that to appear in the phase if one works in the Fermi frame (see~(45) in~\cite{Li&Ni}).
But, there is an extra contribution in the non-inertial~frame, that may be interpreted
as a time-dependent correction to~\eqref{eq:phase-v1}.~It~becomes~non-negligible
with respect to~\eqref{eq:phase-v1} if $t \gtrsim 16.4\,\text{min}$, whereas a
characteristic curvature time at the surface~of Earth is roughly given by
$(r_\oplus/c)\sqrt{r_\oplus/r_{S,\,\oplus}} \approx 9.5\,\text{min}$.~It is unclear
if it is~feasible~to~design~an experiment that
could test this non-inertial-frame contribution to the phase.

The next-to-leading-order curvature correction to the locally Minkowski wave
packet~\eqref{eq:app-wp-0} vanishes in vacuum on the classical geodesic.
Hence, this provides a sub-leading correction to the gravitational tidal effect
we have considered above. Nevertheless, we find no contribution to the
classical geodesic in the non-relativistic limit in Schwarzschild spacetime,
that depends on the first (covariant) derivative~of the Riemann tensor, i.e.
\beqa\label{eq:rnc-trajectory-3}
\langle x^a \rangle_{(3)} & \equiv & \sum\limits_{n \,=\, 0}^{\infty}\delta_{(3)}^{(n)} x^a\,,
\eeqa
where, up to the fourth order in the metric derivatives,
\bsubeqs
\beqa
\delta_{(3)}^{(0)}x^a & \xrightarrow[V \,\rightarrow\, 0]{} & 0\,,
\\[2mm]
\delta_{(3)}^{(1)}x^a & \xrightarrow[V \,\rightarrow\, 0]{} & 0\,.
\eeqa
\esubeqs
Note, for the same reason as in the case of the leading-order correction, there is no
curvature contribution to the normalisation factor~\eqref{eq:normalisation-factor-0} at NLO.

It is straightforward to compute higher-order corrections
to~\eqref{eq:newton-trajectory-2}. These corrections yield
\beqa\label{eq:newton-trajectory-3}
\frac{d^2}{d\tau^2}\,\langle x^i(\tau) \rangle &\approx& - g_\oplus
\left(1+\frac{D^2}{M^2c^2}- 
\frac{r_{S,\,\oplus}\hbar^2}{48(r_\oplus)^3}\left(\frac{16}{D^2}-\frac{91}{M^2c^2}\right)\right)
\delta_3^i\,,
\eeqa
where we only consider $\tau^2$-dependent terms in
$\langle x^i(\tau) \rangle$,~because higher-order polynomials~in~$\tau$
contribute to higher-order terms in the classical trajectory $x^i(\tau)$.
The wave-packet centre of mass falls down with a constant acceleration
which depends on the Lagrangian mass~$M$,~the momentum variance $D$ and
higher-order metric derivatives. This is our main result.

\subsubsection{NNLO curvature correction}
\label{sec:nnlo}

We find eighteen curvature-dependent terms in the next-to-next-to-leading order (NNLO):
\beqa\label{eq:wps-rc-lot-3}
\phi_{Y,K}^{(4)}(y) &=& e^{-iK{\cdot}y}\big(
R_{acbd;ef}K^aK^b I^{cdef} +
R_{acge}R_{bdhf}K^aJ^{bcdefgh}\big)\,,
\eeqa
where by definition
\bsubeqs\label{eq:wps-rc-lot-3-I}
\beqa
I^{cdef} &\equiv& f_1y^cy^dy^ey^f + f_2y^cy^dK^ey^f +
f_3y^cy^dy^eK^f + f_4y^cy^dK^eK^f + f_5y^cy^d\eta^{ef}\,,
\\[3mm]
J^{bcdefgh} &\equiv& f_6K^b y^cy^dy^ey^fK^gK^h
+ K^b\big(f_7y^cy^dy^ey^f + f_{8}y^cy^dK^ey^f + f_{9}y^cy^dK^eK^f\big)\,\eta^{gh}
\nonumber\\[1mm]&&
+\,\big(f_{10}y^{b}y^ey^f + f_{11}K^{b}y^ey^f
+ f_{12}y^bK^ey^f + f_{13}y^by^eK^f + f_{14}K^bK^ey^f
\nonumber\\[1mm]&&
+\,f_{15}y^bK^eK^f + f_{16}y^b\eta^{ef} +f_{17}K^b\eta^{ef}
+ f_{18}K^bK^eK^f\big)\,\eta^{gh}\eta^{cd}\,.
\eeqa
\esubeqs

Substituting~\eqref{eq:wps-rc-lot-3} with~\eqref{eq:wps-rc-lot-3-I}
into~\eqref{eq:wps-rc-lot-n} with $n = 4$, we obtain
\beqa\label{eq:wps-rc-lot-3-f-i}
\partial{\cdot}\partial f_i - 2iK{\cdot}\partial f_i &=& F_i\,,
\eeqa
where the first batch of the $F$-functions reads
\bsubeqs
\beqa
F_1 &=& \frac{1}{20} - 16\dot{f}_1\,,
\\[1mm]
F_2 &=& 2if_1 - 2f_1^\prime - 12\dot{f}_2\,,
\\[2mm]
F_3 &=& 2if_1 - 2f_1^\prime - 12\dot{f}_3\,,
\\[2mm]
F_4 &=& 2if_2 - 2f_2^\prime + 2if_3 - 2f_3^\prime-8\dot{f}_4\,,
\\[2mm]
F_5 &=& -2f_1-8\dot{f}_5\,,
\\[1mm]
F_6 &=& - \frac{1}{3}\,(v_1^{\prime\prime}-2iv_1^\prime-v_1)-16\dot{f}_6\,,
\eeqa
\esubeqs
where we remind that the dot and prime stand for the derivative with respect to $y^2$
and~$K{\cdot}y$, respectively. Taking into account, first, dimensions of the $f$-functions
and, second, assuming that the curvature-dependent factor in~\eqref{eq:wps-rc-lot-3}
is a polynomial of maximal degree six both in~$y$ and in $K$, as this is the case in
de-Sitter spacetime for the solution derived in~\cite{Emelyanov-2020}, we obtain a
solution for the first batch of the $f$-functions, which generically depends on twelve 
constant parameters, assuming that $v_1$ is given
by~\eqref{eq:wps-rc-lot-1-v-sol}.~Next, the second batch of the~$F$-functions~can be treated, namely
\bsubeqs
\beqa
F_7 &=& \frac{1}{15}-16\dot{f}_7\,,
\\[1mm]
F_{8} &=& 4if_7 - 4f_7^\prime + \frac{4}{3}\,(v_1^\prime-iv_1) -12\dot{f}_{8}\,,
\\[2mm]
F_{9} &=& 2if_{8} - 2f_{8}^\prime - 8f_6 - 8\dot{f}_{9}\,,
\\[2mm]
F_{10} &=& -\frac{4i}{45}-12\dot{f}_{10}\,,
\\[1mm]
F_{11} &=& 2if_{10}-2f_{10}^\prime + 4f_1 - 10f_7-8\dot{f}_{11}\,,
\\[2mm]
F_{12} &=& 2if_{10}-2f_{10}^\prime + 4f_1 - \frac{2}{3}\,v_1- 8\dot{f}_{12}\,,
\eeqa
\esubeqs
giving the second batch of the $f$-functions to depend on extra eleven constants.
And, finally, we have
\bsubeqs
\beqa
F_{13} &=& 2if_{10}-2f_{10}^\prime - 8f_1+ 8f_7-8\dot{f}_{13}\,,
\\[2mm]
F_{14}  &=& 4if_{11}-4f_{11}^\prime+2if_{12}-2f_{12}^\prime+2if_{13}-2f_{13}^\prime
-4f_{8} -4\dot{f}_{14}\,,
\\[2mm]
F_{15} &=& 2if_{12}-2f_{12}^\prime+2if_{13}-2f_{13}^\prime+2f_{8}-4\dot{f}_{15}\,,
\\[2mm]
F_{16} &=& -3f_{10} - 4\dot{f}_{16}\,,
\\[2mm]
F_{17} &=& -2f_{11} - f_{13} + 2if_{16}-2f_{16}^\prime\,,
\\[2mm]
F_{18} &=& 2if_{14}-2f_{14}^\prime+ 2if_{15}-2f_{15}^\prime-2f_{9}\,.
\eeqa
\esubeqs
The $f$-functions that solve these equations depend on additional twelve constant
parameters. In total,~\eqref{eq:wps-rc-lot-3} depends on 35 dimensionless parameters
which need to be determined.~It is~not clear to us how these can be unambiguously
achieved. For this reason, it is impossible~at~this stage to make any predictions
for the phase shift at NNLO.

At this order, the curvature-dependent correction does not vanish on the classical~geodesic.
This is the main reason why we consider this order. One of the consequences of this property
is that the Wightman two-point function, defined as
\beqa
W(x,X) &\equiv& \lim\limits_{D\,\rightarrow\, \infty} \frac{D^2}{4\pi^2N}\,\varphi_{X,P}(x)\,,
\eeqa
acquires a non-trivial contribution which can be brought to the form obtained
in~\cite{Bunch&Parker,Christensen} by fixing two (at least in Schwarzschild spacetime)
of all free parameters. Another consequence is the fact that the wave-packet normalisation
factor depends now on the Riemann tensor~at $x = X$. Specifically, we have
from~\eqref{eq:quantum-geodesic-curvature-expansion} that
\beqa\label{eq:rnc-trajectory-4}
\langle x^a \rangle_{(4)} & \equiv & \sum\limits_{n \,=\, 0}^{\infty}\delta_{(4)}^{(n)} x^a\,,
\eeqa
where, up to the fourth order in the metric derivatives, we obtain in Schwarzschild spacetime
that
\beqa
\delta_{(4)}^{(0)}x^a & \xrightarrow[V \,\rightarrow\, 0]{} &
+\,\frac{(g_{\oplus})^2\hbar^4}{D^4c^4(r_\oplus)^2}
\sum\limits_{n \,=\,-1}^\infty\,C_n\Big(\frac{D}{Mc}\Big)^{2n}\delta_0^a\,c\tau\,,
\eeqa
where $C_n$ are real and depend on the free parameters. Therefore, the normalisation factor
\beqa
N & \xrightarrow[V \,\rightarrow\, 0]{} & N^{(0)}\left(1 - \frac{(r_{S,\,\oplus})^2\hbar^4}{8D^4(r_\oplus)^6}
\sum\limits_{n \,=\,-1}^\infty\,C_n\Big(\frac{D}{Mc}\Big)^{2n}\right)
\eeqa
depends on the curvature tensor at the point $X$.

\section{Concluding remarks}

The main goal of this article was to study free fall in quantum physics.~In order to~address
this question, however, a quantum-particle model in curved space is required.~The~conceptual
idea we have put forward in this regard is to covariantly generalise asymptotic particle
states to underlie $S$-matrix in elementary particle physics. First, this approach respects
the general principle of relativity in the sense that the quantum-particle state $|\varphi_{X,P}\rangle$
unitarily transforms under the diffeomorphism group by construction, i.e.
\beqa
|\varphi_{X,P}\rangle &\rightarrow& |\varphi_{\tilde{X},\tilde{P}}\rangle
\;=\; \hat{U}\,|\varphi_{X,P}\rangle\,,
\eeqa
where $\hat{U}$ is a unitary operator that is related to the coordinate transformation
$x \rightarrow \tilde{x} = \tilde{x}(x)$, and, thereby, $\tilde{X} = \tilde{x}(X)$
and $\tilde{P}(\tilde{X})$ is a covariant vector obtained from $P(X)$ by means of~the
tensorial transformation rule. Second, locally and at any point $X$, the wave packet
$\varphi_{X,P}(x)$~to be associated with $|\varphi_{X,P}\rangle$ is represented in a
local inertial frame through the superposition~of Minkowski plane waves, implying that
$|\varphi_{X,P}\rangle$ is related to one of the unitary and irreducible representations
of the Poincar\'{e} group. This implements~the~Einstein~equivalence~principle~at
quantum level.~In other words, quantum-field-theory techniques applied for the
description~of microscopic processes in collider physics are reliable in any
space-time region of the Universe, assuming that a local curvature length 
there is much bigger than a characteristic length~scale to describe a given quantum system
(see~\cite{Emelyanov&Klinkhamer} for an application of this approach).

It is worth emphasising that we have focused here on the construction of
diffeomorphism-invariant quantum states,~$|\varphi_{X,P}\rangle$. In the semi-classical limit
(see below), these states describe particles (approximately) moving along geodesics. Each of
these geodesics is determined by its tangent vector $P/M$ at $X$. In accordance with Heisenberg's
uncertainty principle,~$|\varphi_{X,P}\rangle$ are (essentially) localised in a certain
space volume at a given moment of time (see Sec.~II.3.1 in~\cite{Haag} for further details). Such
localised states can thus be employed to obtain pseudo-local diffeomorphism-invariant observables
in gravity, proposed in~\cite{Giddings&Marolf&Hartle}.

A few experiments have been performed up to now, that require for their explanation~both
quantum theory and gravity. The elementary-particle model we have proposed is, first, in
full agreement with the quantum interference induced by the Earth's gravitational
field~\cite{Colella&Overhauser&Werner}.~This is a non-inertial-frame effect which is
accordingly absent in local inertial frames. It was~later experimentally proved that this
effect also exists in accelerated frames~\cite{Bonse&Wroblewski}.~Loosely speaking, 
these experiments show that the quantum interference cannot be used to distinguish
between uniform gravity and acceleration~\cite{Nauenberg}. The wave function
$\varphi_{X,P}(x)$ is naturally consistent with these observations, as shown in
Sec.~\ref{sec:wpp}. Second, there has been recently observed a phase shift
due to the space-time curvature~\cite{Asenbaum&etal}.~A theoretical basis for this
experimental result~was established in~\cite{Audretsch&Marzlin-1994-I}, assuming
a stationary spacetime. The wave-function solution $\varphi_{X,P}(x)$~we have
obtained in this article agrees with~\cite{Audretsch&Marzlin-1994-I} and, consequently,
with the observation~\cite{Asenbaum&etal},~yet this provides a generic result which is
applicable in vacuum in non-stationary spacetimes~as well (see also~\cite{Emelyanov-2020}
for a non-perturbative (in curvature) result in de-Sitter spacetime).

The covariant wave packet $\varphi_{X,P}(x)$ is therefore suitable for studying free fall of
quantum matter. The main result of this article is the observation that the wave-packet
centre~of~mass falls down with the acceleration that depends on both the Lagrangian mass
$M$ and the wave-packet momentum variance $D$, which is given in~\eqref{eq:newton-trajectory-3}.
It implies that free fall is not universal at quantum level. However, it appears hard to predict
how accurate an experimental test~of free fall must be to observer this effect.~In fact, this
prediction depends on $D$~which~essentially defines the initial quantum-particle state. To
pass experimental constraints on the (possible) violation of the universality of free fall, the
momentum variance $D$ must necessarily be much bigger than
$(\hbar/r_\oplus)\sqrt{r_{S,\,\oplus}/r_\oplus} \sim 10^{-18}\,\text{eV/c}$ and
much smaller than $Mc$. To put it differently, the characteristic (initial) extent of the wave
packet must be much bigger than the Compton wavelength $\hbar/Mc$ and much smaller than
the local curvature length $r_\oplus\sqrt{r_\oplus/r_{S,\,\oplus}} \sim 10^{11}\,\text{m}$.
This qualitatively agrees with the results of~\cite{Emelyanov-2020}
(see also appendices in~\cite{Roura-2020}).

The non-universality of free fall implies that the weak equivalence principle cannot hold~in
quantum theory, if conceived as the free-fall universality~\cite{Casola&Liberati&Sonego}.~There
are many other arguments supporting this assertion, see~\cite{Laemmerzahl}.~For~this reason,
this reference proposed a quantum version of the weak equivalence principle. If adapted to the
position expectation value, this quantum principle basically
states that this expectation value cannot depend on $M$. The lower bound on $\hbar/D$
arises from $-\frac{1}{2}\Gamma_{ij}^a(X)\langle y^i y^j \rangle_{(0)}$ to give a quantum
correction to $-\frac{1}{2}\Gamma_{00}^a(X)(y^0)^2$. The expectation value 
$\langle y^i y^j \rangle_{(0)}$ characterises the wave-packet extent which is subject
to spreading. It is a universal quantum phenomenon, see~\cite{Merzbacher}, which is
described by $(D/M)^2(y^0)^2$.~From~an experimental point of view, this is the most relevant 
source of the deviation from the classical law of free fall, as the upper bound on
$\hbar/D$ is violated for a wave packet of size comparable~to or bigger than the Earth-orbit
radius.

For an experiment aiming at testing the weak equivalence
principle by making use of a pair of atoms, one may estimate $\hbar/D$ by
an atomic diameter.~Defining the E\"{o}tv\"{o}s parameter~as
\beqa
\eta(A,B) &\equiv& 2\,\frac{g_A-g_B}{g_A+g_B}\,,
\eeqa
where $g_{A}$ and $g_{B}$ stand for free-fall accelerations of two atoms $A$ and $B$,
respectively, we get
\beqa\label{eq:estimate-1}
\eta(\text{Au},\text{Al}) &\approx& -8.88{\times}10^{-16}\,.
\eeqa
This cannot be compared to the experimental upper bound of the order of $10^{-11}$
found~in~\cite{Roll&Krotkov&Dicke}, as~aluminum and gold cylinders used in the
experiment give the E\"{o}tv\"{o}s parameter $\eta \sim 10^{-79}$. The theoretical
estimate~\eqref{eq:estimate-1} is close to the precision of the
MICROSCOPE experiment~\cite{Touboul&etal}, where, however, a relative acceleration
of alloy masses,~rather than~atoms,~has~been~probed~on board of a satellite.
Next, if we consider ${}^{87}\text{Rb}$ and
${}^{39}\text{K}$ atoms, then the E\"{o}tv\"{o}s parameter~is dominated in this pair by
potassium atoms, as these are lighter and smaller. In this case,~we obtain the following estimate:
\beqa\label{eq:estimate-2}
\eta(\text{Rb},\text{K}) &\approx& -1.42{\times}10^{-16}\,.
\eeqa
This is by $9$ orders of magnitude smaller than the sensitivity of a matter-wave
interferometer used to test the free-fall universality with rubidium and potassium
atoms~\cite{Schlippert&etal,Albers&etal}, whereas by $4$ orders of magnitude smaller
than the atom-interferometer sensitivity (employing ${}^{85}\text{Rb}$~and ${}^{87}\text{Rb}$
isotopes) achieved in~\cite{Asenbaum&etal-2020}. A novel measurement
technique~\cite{Roura-2017} allowed herein to reduce systematic
errors~\cite{Amico&etal}, making atom interferometers promising for quantum tests of
the weak equivalence principle with even higher sensitivity.

The main result~\eqref{eq:newton-trajectory-3} relies on an assumption that the position
expectation value is given~in the Schwarzschild frame by the
integral~\eqref{eq:quantum-geodesic}. This definition tacitly uses the
constant-proper-time Cauchy surfaces, as $P^i = 0$ has been set in the integral
computations. Still,~the~position expectation value depends on a
Cauchy-surface choice~\cite{Emelyanov-2020}. Up to the first order in derivatives of the metric
tensor, we find
\beqa\label{eq:cf-2}
\frac{d^2}{dt^2}\,\langle x^i(t) \rangle &\approx&
- g_\oplus\left(1-\frac{D^2}{M^2c^2}\right)\delta_3^i\,,
\eeqa
where Cauchy surfaces of constant $t$ have been considered.~Note that the Schwarzschild
time~$t$ and proper time $\tau$ (over the geodesic~\eqref{eq:classical-geodesic} with
$P^i = 0$) roughly equal if $t \ll c/g_\oplus \sim 10^7\,\text{s}$.~At this order of the approximation,
the difference between~\eqref{eq:cf-2} and~\eqref{eq:cf-1} effectively stems from replacing
$\partial_0$ by $\partial_0 - g_\oplus y^0\partial_3$. Therefore, quantum tests of the universality
of free fall might also show if quantum particles measure proper time.~It would then provide
another witness for the proper-time role in quantum theory, see~\cite{Zych&etal-2011}.~A theoretical argument in favor of that could be based on the circumstance that, in the Riemann
normal frame, $\langle y^a \rangle$ exactly~equals~$(P^a/M)\,\tau$, independent
on the ratio $D/Mc$ (at least up to the order studied), when the Cauchy surfaces
of constant proper time are treated. Indeed, this choice ensures that the curvature-dependent 
normalisation factor
$N$ does not modify that equality.~In this case and in the Riemann~normal frame,
the wave-function centre of mass propagates over the classical
geodesic~\eqref{eq:classical-geodesic} as if the wave packet were structureless.

The expectation value $\langle x^a \rangle$ makes no use of
position operator. It is not even clear how to self-consistently define quantum-mechanics
operators with corresponding expectation values in curved spacetime. Still, such quantities
as energy and three-momentum of the wave packet can be established without introducing the
quantum-mechanics operator formalism~by~solely employing the quantum-field algebra:
\beqa\label{eq:momentum}
\langle p_a \rangle &\equiv& {\int_\Sigma}d\Sigma^b(y)\,\big(
\langle \varphi_{Y,P}|\hat{T}_{ab}(y)|\varphi_{Y,P}\rangle - 
\langle \Omega|\hat{T}_{ab}(y)|\Omega\rangle
\big)\,,
\eeqa
where the scalar-field stress tensor reads
\beqa
\hat{T}_{ab}(y) &\equiv& \nabla_a\hat{\Phi}(y)\nabla_b\hat{\Phi}(y) - \frac{1}{2}\,g_{ab}(y)
\Big(\nabla_c\hat{\Phi}(y)\nabla^c\hat{\Phi}(y) - M^2\hat{\Phi}^2(y)\Big)
\nonumber\\[1mm]
&&+\,\frac{1}{6}\,\Big(G_{ab}(y) - \nabla_a\nabla_b + g_{ab}(y)\nabla_c\nabla^c\Big)\hat{\Phi}^2(y)
\eeqa
for the scalar-field model studied, where $G_{ab}(y)$ is the Einstein tensor. The
integrand in~\eqref{eq:momentum} can be expressed through $\varphi_{Y,P}(y)$ and
its complex conjugate by means of the quantum-field algebra, wherein
$-\langle \Omega|\hat{T}_{ab}(y)|\Omega\rangle$ serves to set the quantum-vacuum
four-momentum to zero. If the Riemann tensor is now neglected, $\langle p^i \rangle$
approaches $M\partial_0\langle y^i \rangle$ in the non-relativistic limit, while
$\langle p^0 \rangle$ naturally contains that what would be the quantum-mechanical
expectation value of the three-momentum-squared operator. It remains to understand how to
obtain $\langle y^a p_b \rangle$~and $\langle p_b y^a \rangle$, which might give a
curvature-dependent commutation relation.

The quantum state $|\varphi_{X,P}\rangle$ cannot yet be treated as a
satisfactory model for an elementary particle. Indeed, this state does not
source gravity, i.e. $\langle\varphi_{X,P}|\hat{h}_{\mu\nu}(x)|\varphi_{X,P}\rangle = 0$,
where $\hat{h}_{\mu\nu}(x)$ is the graviton field, although
$\langle\varphi_{X,P}|\hat{T}_{\mu\nu}(x)|\varphi_{X,P}\rangle \neq
\langle \Omega|\hat{T}_{\mu\nu}(x)|\Omega\rangle$. A proper generalisation~of
the operator $\hat{a}^\dagger(\varphi_{X,P})$ given in~\eqref{eq:creation-operator}
is required to go beyond the test-particle approximation. For this reason, it is unclear
whether~\eqref{eq:newton-trajectory-3} implies that the gravitational mass differs
from~the inertial mass and how these two masses are related to the Lagrangian mass
$M$ in gravity.

Another aspect of the problem is how to determine the free dimensionless parameters~that
enter the wave function $\varphi_{X,P}(x)$.~It may require study of this
problem in other spacetimes,~by searching for a unique form of $\varphi_{X,P}(x)$
reducing to preferred wave functions in corresponding curved spacetimes. Besides, it may
require quantum gravity for dealing with this question.~In particular, it might be feasible at least to
re-produce our results by using the effective-theory approach to quantum gravity, by setting up
no background gravitational field and~replacing the Earth by a heavy quantum particle.~This
approach has already
provided quantum-gravity corrections to hyperbolic-like trajectories~\cite{Bjerrum-Bohr&etal-2015-1,Donoghue&El-Menoufi,Bjerrum-Bohr&etal-2015-2}: In quantum gravity,
such trajectories~cannot be solutions of the geodesic
equation.

Finally, the quantum-mechanical phase is given in the semi-classical limit by $+iS(x,X)/\hbar$,
where $S(x,X) = -Mc^2\tau(x,X)$ is the single-particle (on-shell) action. Curvature-dependent
corrections in $\varphi_{X,P}(x)$ may in general contribute to this phase on a classical
geodesic.~In~fact, the leading correction in vacuum is proportional to the curvature tensor
squared, as shown in Sec.~\ref{sec:nnlo}. This means that the classical action for a single
particle in curved spacetime~has~a limited application in quantum theory.

\section*{
ACKNOWLEDGMENTS}

I am thankful to Daniel Terno for the reference~\cite{Brodutch&etal}.~It is also my pleasure to
thank~Andrzej Czarnecki
and Albert Roura for the references~\cite{Czrnecka&Czarnecki}
and~\cite{Roura-2020,Asenbaum&etal-2020,Roura-2017}, respectively.


\begin{thebibliography}{99}

\bibitem{Einstein-1905}
A. Einstein,
Annalen der Physik {\bf 322} (1905) 891.

\bibitem{Wigner-1939}
E. Wigner,
Annals of Math. {\bf 40} (1939) 149.

\bibitem{McReynolds}
A.W. McReynolds,
Phys. Rev. {\bf 83} (1951) 172.

\bibitem{Dabbs&Harvey&Paya&Horstmann}
J.W.T. Dabbs, J.A. Harvey, D. Paya, H. Horstmann,
Phys. Rev. {\bf 139} (1965) B756.

\bibitem{Koester}
L. Koester,
Phys. Rev. D{\bf 14} (1976) 907.

\bibitem{Einstein-1916}
A. Einstein,
Annalen der Physik {\bf 354} (1916) 769.

\bibitem{Emelyanov-2020}
V.A. Emelyanov, Eur. Phys. J. C {\bf 81} (2021) 189.

\bibitem{Casola&Liberati&Sonego}
E. Di Casola, S. Liberati, S. Sonego, 
Am. J. Phys. {\bf 83} (2015) 39.

\bibitem{Lehmann&Symanzik&Zimmermann}
H. Lehmann, K. Symanzik, W. Zimmermann, Nuovo Cimento {\bf 1} (1955) 205.

\bibitem{DeWitt-2003}
B.S. DeWitt,
\emph{The Global Approach to Quantum Field Theory} (Oxford University Press, 2003).

\bibitem{DeWitt-1975}
B.S. DeWitt, Phys. Rep. {\bf 19} (1975) 295.

\bibitem{Weinberg}
S. Weinberg,
\emph{Quantum Theory of Fields}
(Cambridge University Press, 1995).

\bibitem{Mukhanov}
V.F. Mukhanov,
\emph{Physical Foundations of Cosmology}
(Cambridge University Press, 2005).

\bibitem{Donoghue}
J.F. Donoghue, Phys. Rev. D {\bf 50} (1994) 3874.

\bibitem{Petrov}
A.Z. Petrov,
\emph{Einstein Spaces}
(Pergamon Press Ltd., 1969).

\bibitem{Naumov&Naumov}
D.V. Naumov, V.A. Naumov,
J. Phys. G: Nucl. Part. Phys. {\bf 37} (2010) 105014.
 
 \bibitem{Naumov}
D.V. Naumov,
Phys. Part. Nuclei Lett. {\bf 10} (2013) 642.

\bibitem{DeWitt-1965}
B.S. DeWitt,
\emph{Dynamical Theory of Groups and Fields} (Gordon and Breach, 1965).

\bibitem{Merzbacher}
E. Merzbacher,
\emph{Quantum Mechanics} (3rd Edition, John Wiley \& Sons, Inc., 1998).

\bibitem{Colella&Overhauser&Werner}
R. Colella, A.W. Overhauser, S.A. Werner, 
Phys. Rev. Lett. {\bf 34} (1975) 1472.

\bibitem{Colella&Overhauser}
R. Colella, A.W. Overhauser, 
Phys. Rev. Lett. {\bf 33} (1974) 1237.

\bibitem{Bonse&Wroblewski}
U. Bonse, T. Wroblewski,
Phys. Rev. Lett. {\bf 51} (1983) 1401.

\bibitem{Nauenberg}
M. Nauenberg,
Am. J. Phys. {\bf 84} (2016) 879.

\bibitem{Luschikov&Frank}
V.I. Luschikov, A.I. Frank, JETP Lett. {\bf 28} (1978) 559.

\bibitem{Nesvizhevsky&etal}
V.V. Nesvizhevsky \emph{et al.}, Nature {\bf 415} (2002) 297.

\bibitem{Mannheim}
P.D. Mannheim,
Phys. Rev. A {\bf 57} (1998) 1260.

\bibitem{Zych&etal-2021}
M. Zych \emph{et al.},
Class. Quantum Grav. {\bf 29} (2012) 224010.

\bibitem{Rideout&etal-2021}
D. Rideout \emph{et al.},
Class. Quantum Grav. {\bf 29} (2012) 224011.

\bibitem{Brodutch&etal}
A. Brodutch \emph{et al.},
Phys. Rev. D {\bf 91} (2015) 064041.

\bibitem{Czrnecka&Czarnecki}
A.P. Czarnecka, A. Czarnecki,
Am. J. Phys. {\bf 89} (2021) 634.

\bibitem{Asenbaum&etal}
P. Asenbaum \emph{et al.},
Phys. Rev. Lett. {\bf 118} (2017) 183602.

\bibitem{Anandan}
J. Anandan,
Phys. Rev. D {\bf 30} (1984) 1615.

\bibitem{Audretsch&Marzlin-1994-I}
J. Audretsch, K.-P. Marzlin,
Phys. Rev. A {\bf 50} (1994) 2080.

\bibitem{Audretsch&Marzlin-1994-II}
J. Audretsch, K.-P. Marzlin,
J. Phys. II France {\bf 4} (1994) 2073.

\bibitem{Audretsch&Marzlin-1996}
J. Audretsch, K.-P. Marzlin,
Phys. Rev. A {\bf 53} (1996) 312.

\bibitem{Bongs&etal}
K. Bongs \emph{et al.},
Appl. Phys. B {\bf 84} (2006) 599.

\bibitem{Li&Ni}
W.-Q. Li, W.-T. Ni,
J. Math. Phys. {\bf 20} (1979) 1925.

\bibitem{Marletto&Vedral}
C. Marletto, V. Vedral,
Front. Phys. {\bf 8} (2020) 176.

\bibitem{Bunch&Parker}
T.S. Bunch, L. Parker,
Phys. Rev. D {\bf 20} (1979) 2499.

\bibitem{Viola&Onofrio}
L. Viola, R. Onofrio,
Phys. Rev. D {\bf 55} (1997) 455.

\bibitem{Christensen}
S.M. Christensen, 
Phys. Rev. D {\bf 14} (1976) 2490.

\bibitem{Emelyanov&Klinkhamer}
V.A. Emelyanov, F.R. Klinkhamer,
Acta Phys. Polon. B {\bf 52} (2021) 805.

\bibitem{Haag}
R. Haag,
\emph{Local Quantum Physics. Fields, Particles, Algebras} (Springer-Verlag, 1996).

\bibitem{Giddings&Marolf&Hartle}
S.B. Giddings, D. Marolf, J.B. Hartle,
Phys. Rev. D {\bf 74} (2006) 064018.

\bibitem{Roura-2020}
A. Roura,
Phys. Rev. X {\bf 10} (2020) 021014.

\bibitem{Laemmerzahl}
C. L\"{a}mmerzahl,
Gen. Rel. Grav. {\bf 28} (1996) 1043.

\bibitem{Roll&Krotkov&Dicke}
P.G. Roll, R. Krotkov, R.H. Dicke,
Ann. Phys. {\bf 26} (1964) 442.

\bibitem{Touboul&etal}
P. Touboul \emph{et al.},
Phys. Rev. Lett. {\bf 119} (2017) 231101.

\bibitem{Schlippert&etal}
D. Schlippert \emph{et al.},
Phys. Rev. Lett. {\bf 112} (2014) 203002.

\bibitem{Albers&etal}
H. Albers \emph{et al.},
Eur. Phys. J. D {\bf 74} (2020) 145.

\bibitem{Asenbaum&etal-2020}
P. Asenbaum \emph{et al.},
Phys. Rev. Lett. {\bf 125} (2020) 191101.

\bibitem{Roura-2017}
A. Roura,
Phys. Rev. Lett. {\bf 118} (2017) 160401.

\bibitem{Amico&etal}
G. D'Amico \emph{et al.},
Phys. Rev. Lett. {\bf 119} (2017) 253201.

\bibitem{Zych&etal-2011}
M. Zych \emph{et al.},
Nat. Commun.  {\bf 8} (2011) 505.

\bibitem{Bjerrum-Bohr&etal-2015-1}
N.E.J. Bjerrum-Bohr \emph{et al.},
Phys. Rev. Lett. {\bf 114} (2015) 061301.

\bibitem{Donoghue&El-Menoufi}
J.F. Donoghue, B.K. El-Menoufi,
J. High Energy Phys. {\bf 05} (2015) 118.

\bibitem{Bjerrum-Bohr&etal-2015-2}
N.E.J. Bjerrum-Bohr \emph{et al.},
Int. J. Mod. Phys. D {\bf 24} (2015) 1544013.
\end{thebibliography}
\end{document}